\documentclass[twocolumn,aps,pre,floats,superscriptaddress,showpacs]{revtex4}
\usepackage{amsmath}
\usepackage{graphics}
\usepackage{epstopdf}
\usepackage{epsfig}

\begin{document}

\author{D. Soriano-Pa\~nos}
\affiliation{Departamento de F\'isica de la Materia Condensada, Universidad de Zaragoza, 50009 Zaragoza, Spain}
\affiliation{GOTHAM lab, Instituto de Biocomputaci\'on y F\'isica de Sistemas Complejos BIFI, Universidad de Zaragoza, 50018 Zaragoza, Spain}
\author{H. Arias-Castro}
\affiliation{Departamento de Matem\'aticas, Universidad del Valle, 25360 Santiago de Cali, Colombia}
\author{F. Naranjo-Mayorga}
\affiliation{Departamento de F\'{\i}­sica, Universidad Pedag\'ogica y Tecnol\'ogica de Colombia, 150003 Tunja, Colombia} 
\author{J. G\'omez-Garde\~nes}
\affiliation{Departamento de F\'isica de la Materia Condensada, Universidad de Zaragoza, 50009 Zaragoza, Spain}
\affiliation{GOTHAM lab, Instituto de Biocomputaci\'on y F\'isica de Sistemas Complejos BIFI, Universidad de Zaragoza, 50018 Zaragoza, Spain}

\title{Impact of human-human contagions in the spread of vector-borne diseases.}

\begin{abstract}
This article is aimed at proposing a generalization of the Ross-Macdonald model for the transmission of Vector-borne diseases in which human-to-human contagions are also considered. We first present this generalized model by formulating a mean field theory, checking its validity by comparing to numerical simulations. To make the premises of our model more realistic, we adapt the mean field equations to the case in which human contacts are described by a complex network. In both the mean-field and network-based models, we estimate the value of the epidemic threshold which corresponds to the boundary between the disease-free and epidemic regimes. The expression of this threshold allows us to discuss the impact that human-to-human contagions have on the spread of vector-borne diseases.
\end{abstract}
\maketitle

\section{Introduction}
\label{sec.Intro}

Vector-borne diseases (VBD) are transmitted indirectly among humans by the intermediation of vectors such as mosquitoes, ticks or sand-flies. According to the World Health Organization (WHO), these diseases represent over the 20 $\%$ of all infectious diseases causing more than one billion of infections and one million deaths per year \cite{WHO}. In the recent years there is a growing global concern about VBD since, despite being mainly localized in tropical and ecuatorial areas, these diseases have started to spread across more tempered latitudes due to human impact \cite{climate,N.climate} and the adaptability of the vectors to urban areas. For instance, {\it Aedes albopticus} species, which are responsible for the transmission of some important VBD like Dengue or Chikungunya, can survive in cool temperatures and their eggs had the ability to diapause during winter \cite{Sharp}. These facts have resulted in a wide geographical distribution of  {\it Aedes albopticus}, expanding from South America and Asia to North America and Europe \cite{Albo_global}. As a consequence, in the recent years the first observations of endogenous cases of VBD in the southern part of Europe and the island of Madeira have been reported \cite{Brown,Romi,Madeira}.

A recent example of the present and future threats behind the rapid advance of VBD has been the outbreak of ZIKV epidemics in 2015 and 2016. This disease, originally localized in the pacific area, became a global concern in few months due to its sudden expansion to the Caribbean and South America and, boosted, by the series of neurological abnormalities associated to its infection, mainly newborn microcephaly and the Guillain-Barr\'e syndrome \cite{Demir,Zika,Zika2}. One of the most surprising features of ZIKV is that, unlike most VBD, it can also be transmitted between humans via materna-fetal or sexual transmission \cite{Zikah}. Inspired by this finding, the aim of this work is to study the role that this new contagion pathway plays on the epidemic onset of VBD.

%The advance of VBD demands the study of accurate epidemic models able of capturing their most important ingredients in order to guide the efforts of public health agencies to ameliorate their spreading. Recent studies report that, 

%Therefore, it becomes clear that ZIKV can not be described by the original version of Ross-Macdonald model since it lacks the possibility of having human-to-human contagions. In fact, the existence of this direct infections among humans, demands to incorporte in the study of ZIKV spread many aspects that were recently addressed for usual compartmental models (such as the {\it Susceptible-Infected-Susceptible} or the {\it Susceptible-Infected-Recovered}\cite{Vespignani,}. One of the most important ingredients consist of abandoning the typical mean-field hypothesis of compartmental models to incorporate the ingredient that interactions between humans are dictated by a complex network\cite{Newman,Boccaletti}.

The most usual way of modeling epidemics relies on compartmental models \cite{Anderson,Rohani,Bacaer2011}. Compartmental models consider populations in which agents can adopt a discrete and finite set of states. In the case of VBD, the Ross-Macdonald (RM) framework \cite{Ross,Macdonald,N.education} constitutes the paradigmatic compartmental model for their study. In this model, humans and vectors can be in either of the two possible states: susceptible ($S^H$ for humans and $S^M$ for vectors) or infected ($I^H$ for humans and $I^M$ for vectors). The transitions between these two states occur following different processes. First, contagion events, {\em  i.e.} the transition $S \rightarrow I$, only occur in a crossed way: from infectious vectors (humans) to healthy humans (vectors):
\begin{eqnarray}
I^M + S^H &\xrightarrow{\lambda^{MH}}& I^M + I^H\ , \\ 
I^H + S^M &\xrightarrow{\lambda^{HM}}& I^H + I^M\ , 
\end{eqnarray}
where $\lambda^{HM}$ and $\lambda^{MH}$ are the respective transmission probabilities for each type of contagion. These crossed infections make the RM model a very suitable framework to characterize the evolution of VBD \cite{Smith,Ruan2008,Regan2016,review}. Second, the transitions from Infected to Susceptible are endogenous, {\em  i.e.} they are not the product of any interaction:
\begin{eqnarray}
I^H &\xrightarrow{\mu^{H}}& S^H\ , \\ 
I^M &\xrightarrow{\mu^{M}}& S^M\ , 
\end{eqnarray}
where $\mu^{H}$ accounts for the recovery probability of humans, whereas $\mu^{M}$ is the death rate of vectors. Note that infected vectors are not diseased but they act as simple carriers of the pathogen. Thus, infected vectors are simply replaced at a rate $\mu^{H}$ by the new vectors born as susceptible, {\em i.e.} free of pathogens. Finally, to generalize the RM model by incorporating contagions from human contacts we add the probability $\lambda^{HH}$ that an infected human transmits the disease to a susceptible one:
\begin{equation}
I^H + S^H \xrightarrow{\lambda^{HH}} 2\; I^H\;. 
\end{equation}
The former microscopic transitions between the two compartments for humans and vectors are summarized in Fig.~1.
  
\begin{figure}[t!]
\centering
\includegraphics[scale=0.13]{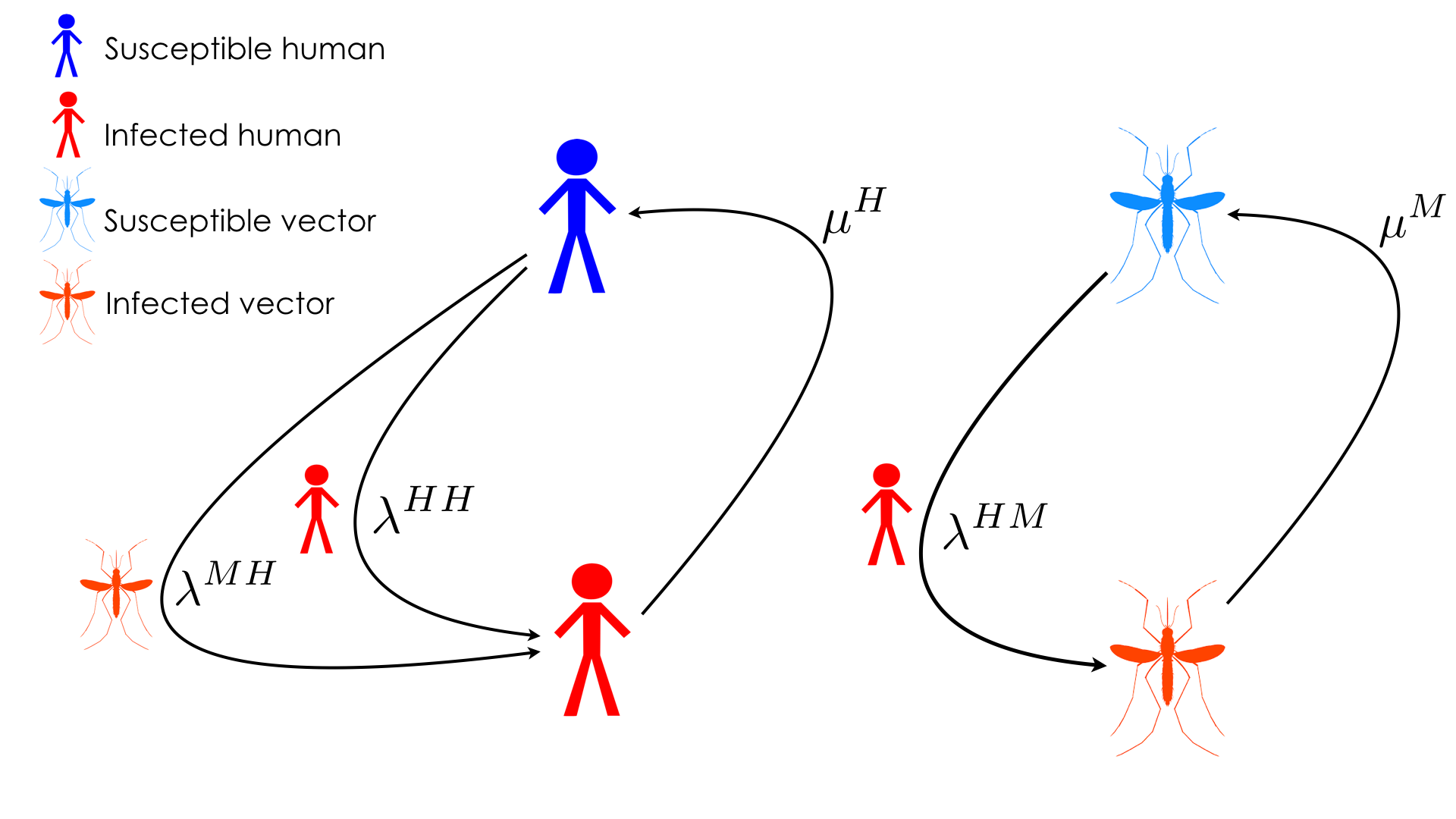}
\caption{Schematic representation of contagion and recovery processes in our modified Ross Macdonald model. The contagion processes are determined by the crossed contagion rates, $\lambda^{MH}$ and $\lambda^{HM}$, and the contagion rate between humans $\lambda^{HH}$. The transition from infected to susceptible are controlled by the human recovery rate, $\mu^H$, and the death rate of vectors, $\mu^M$.}
\end{figure}
  
The structure of this manuscript is as follows. We first adapt the usual mean-field formulation of the RM model to incorporate human-human contagions. We then use the proposed equations to estimate the influence of human-human contagion paths on the epidemic threshold, which is defined as the boundary between the disease-free and the epidemic solutions. Afterwards, to incorporate realistic human interaction patterns, we leave the mean-field assumption by encoding the human contacts in a complex network \cite{Newman,Boccaletti}. To this aim, we will write the dynamical evolution equations by making use of the Microscopic Markov Chain Approach (MMCA) \cite{MMCA1,MMCA2,MMCA3,RMP} and use this formalism to describe the effect of introducing heterogeneous (Scale-free \cite{BA,Newman2}) interaction patterns among humans. To round off, we will derive analytically the epidemic threshold when a general network of human contacts is considered.

\section{Mean field approach}
\label{sec.MF}
\subsection{Model equations}

In order to get a first insight about the effects of adding a new contagion path to the RM model, we propose a mean field theory whose equations are based on discrete Markov chains. Following the original RM model assumptions, we suppose a closed population of $N$ humans and we define $\gamma$ as the ratio between vectors and humans, so that there are $\gamma\cdot N$ vectors. The microscopic (contagion and recovery) processes at work have been already introduced above and summarized in Fig.~1. On the one hand, contagion processes are the outcome of the interactions between vectors and humans and among humans, which are characterized by the contagion probabilities$\{\lambda^{MH},\;\lambda^{HM},\;\lambda^{HH}\}$. To model these interactions, we assume that each vector makes $\beta$ bites per time step  and that each human contacts with other $k$ humans per time step. On the other hand, the recovery processes are determined by the recovery/death rates $\{\mu^{H},\;\mu^{M}\}$.
% Concerning the contagion processes, as we can observe in Figure 1, we suppose that mosquitoes have a biting rate $\beta$. Besides, a susceptible vector, after biting an infected human, can become infected with a probability $\lambda^{HM}$, whereas a susceptible agent can get the disease after contacting with a mosquito with a probability $\lambda^{MH}$. However, the novelty of the proposed model with respect to traditional Ross-Macdonald model is that we assume that each agent interacts with $k$ agents at each time step. Besides, contagion procceses among humans are no longer restricted so that a susceptible agent can become infected after contacting with another infected with a probability $\lambda^{HH}$. Finally, recovery rates are set to $\mu^M$ and $\mu^H$ for mosquitoes and people respectively. 

Following the mean field premises of the original RM model, the state of the whole system is characterized by two variables, $\rho^H(t) $ and $\rho^M(t)$, which indicate us the fraction of infected humans and infected vectors. Taking into account the parameters previously defined, the time evolution of the fraction of infected humans, $\rho^H(t)$, can be expressed as:
\begin{eqnarray}
\rho^{H} (t+1) &=& \left(1-\mu^H\right)\rho^H(t) \nonumber\\
&+& \left(1-\rho^H(t)\right)\left[1-P_{ninf}^{HH}(t)P_{ninf}^{MH}(t)\right]
\label{ec.human}
\end{eqnarray}
where the first term corresponds to the infected agents that do not recover, whereas the second term denotes the susceptible ones who catch the disease. The probabilities of not being infected neither by contact with vectors, $P_{ninf}^{HH}(t)$, nor by contact with infected humans, $P_{ninf}^{MH}(t)$, are given by:
\begin{eqnarray}
P_{ninf}^{HH}(t) &=& \left(1-\lambda^{HH}\rho^{H}\right)^{k}\ ,\\
P_{ninf}^{MH}(t) &=& \left(1-\lambda^{MH}\rho^{M}\right)^{\beta\gamma}\label{MH}\;, 
\end{eqnarray}
where factor $\beta\gamma$ accounts for the number of times that a human is bitten by a vector. Let us note that the proportion between vectors and humans, encoded in parameter $\gamma$, plays a key role on the transmission of the pathogen from vectors to humans, as seen in Eq. (\ref{MH}).

Following the same framework, the time evolution of the fraction of infected vectors, $\rho^M(t)$, can be written as:
\begin{eqnarray}
\rho^M(t+1) &=& \left(1-\mu^M\right)\rho^M(t)\nonumber\\ 
&+& \left(1-\rho^M(t)\right)\left(1-\left(1-\lambda^{HM}\rho^{H}(t)\right)^\beta\right)
\label{ec.mosquito}
\end{eqnarray}
where the first term corresponds to vectors that remain infected, whereas the second term denotes the ones which become infected after biting  infected human. Let us remark that in this case the proportion between vectors and humans does not have any influence since all the vectors are assumed to make $\beta$ bites per time step, regardless of the number of agents in the system under study. 

Summing up, given an initial condition, $\rho^{H}(0)$ and $\rho^{M}(0)$, the iteration of Eqs.(\ref{ec.human})-(\ref{ec.mosquito}) allows us to monitor the time evolution of the spread of a VBD with two contagion mechanisms under a mean-field assumption.
%\begin{eqnarray}
%\rho^M(t+1) &=& \left(1-\mu^M\right)\rho^M(t) + \left(1-\rho^M(t)\right)\left(1-\left(1-\lambda^{HM}\rho^{H}(t)\right)^\beta\right) \ ,\label{ec.human} \\
%\rho^{H} (t+1) &=& \left(1-\mu^H\right)\rho^H(t) + \left(1-\rho^H(t)\right)\left[1-\left(1-\lambda^{MH}\rho^{M}\right)^{\beta\gamma}\left(1-\lambda^{HH}\rho^{H}\right)^{k}\right]\label{ec.mosquito}.
%\end{eqnarray}

\begin{figure}[t!]
\centering
\includegraphics[scale=0.29]{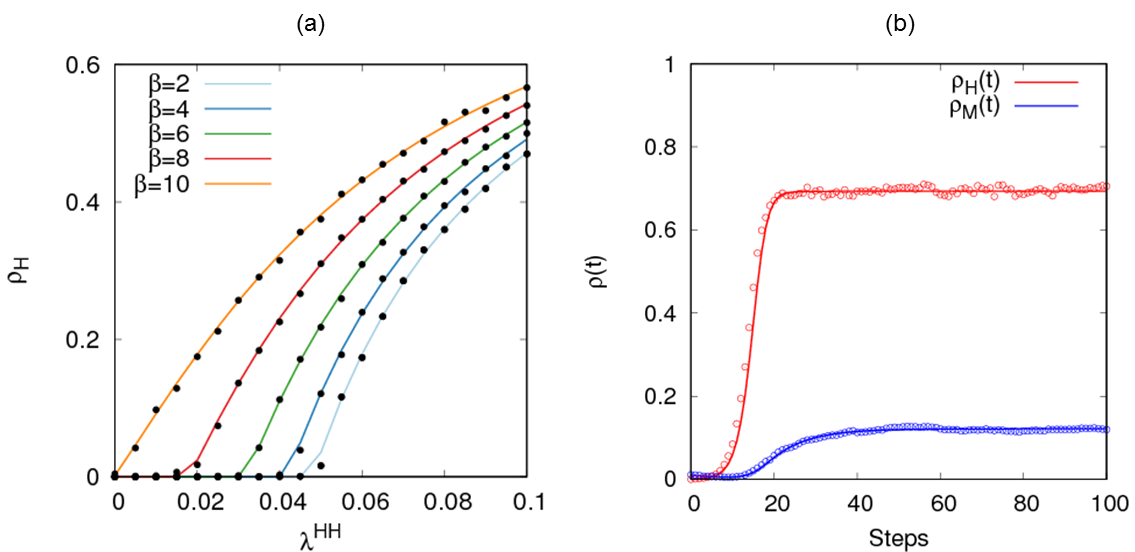}
\caption{Panel (a). Fraction of infected people in the stationary state $\rho^H$ as a function of the human-to-human contagion probability $\lambda^{HH}$. Solid lines correspond to the predictions of our model by integrating Eqs.(\ref{ec.human},\ref{ec.mosquito}) whereas black dots represent the solutions from numerical simulations obtain by averaging $20$ realizations. The color of each line denotes the biting rate $\beta$, while the rest of parameters of our model ($\lambda^{MH},\lambda^{HM},k,\gamma,\mu^H,\mu^M$) are set to ($0.01,0.01,4,2,0.2,0.1$). Panel (b). Temporal evolution of the fraction of infected people (red) and the fraction of infected mosquitoes (blue). Solid lines are obtained by iterating Eqs.~(\ref{ec.human})-(\ref{ec.mosquito}) whereas dots correspond to results from MC simulations. In this case, we have set ($\lambda^{MH},\lambda^{HM},\lambda^{HH},k,\gamma,\beta,\mu^H,\mu^M$) to ($0.01,0.01,0.2,4,2,2,0.2,0.1$).}
\end{figure}

\subsection{Model validation}
In order to validate the former equations, it is customary to compare its predictions about the incidence of a disease, which is defined as the fraction of infected agents in the stationary state, to results from Monte Carlo simulations. These simulation are performed by tracking the state of each vector and human, which changes in time according to the probabilistic rules defined in Fig.~1. This way, we start by infecting a $1\%$ of vectors and then we let the whole system evolve until it reaches the stationary state. Due to the stochastic nature of the microscopic contagion and recovery processes, it is necessary to average the results for several realizations of these Monte Carlo simulations. In turn, theoretical predictions are obtained by iterating Eqs.~(\ref{ec.human})-(\ref{ec.mosquito}) until the stationary solution, ($\rho^H$, $\rho^M$), is reached.

In order to show the accuracy of the mean-field formulation, we have represented in Fig.~2.a the prediction of Eqs.~(\ref{ec.human})-(\ref{ec.mosquito}) about the incidence of a disease as a function of the human-human contagion rate, $\lambda^{HH}$, for several values of the vector biting rates $\beta$ and considering that the population of vectors duplicates that of humans ($\gamma =2$) and a that each human contacts $k=4$ other humans per time step \cite{socialnet}. Regarding Monte Carlo simulations, we have considered a system composed of $N=4000$ humans (and correspondingly $8000$ vectors). The accuracy of Eqs.~(\ref{ec.human})-(\ref{ec.mosquito}) becomes clear from the diagrams $\rho^{H}(\lambda^{HH})$. In addition, it can be observed that increasing the vectors biting rate $\beta$ clearly boosts spreading due to the pronounced decrease of the epidemic threshold, $\lambda^{HH}_c$, here represented as the minimum value of $\lambda^{HH}$ for which $\rho^{H}>0$. To further validate the mean-field formulation we analyze the temporal evolution of the disease incidence. To this aim, in Fig.~2.b, we show (solid lines) the time evolution for the fraction of infected humans, $\rho^{H}(t)$, and vectors, $\rho^{M}(t)$, as obtained by solving (iterating) Eqs.~(\ref{ec.human})-(\ref{ec.mosquito}). The two curves are in perfect agreement with the numerical results obtained from a single run of a Monte Carlo simulation (dots) in a population. 

\begin{figure}[t!]
\centering
\includegraphics[scale=0.27]{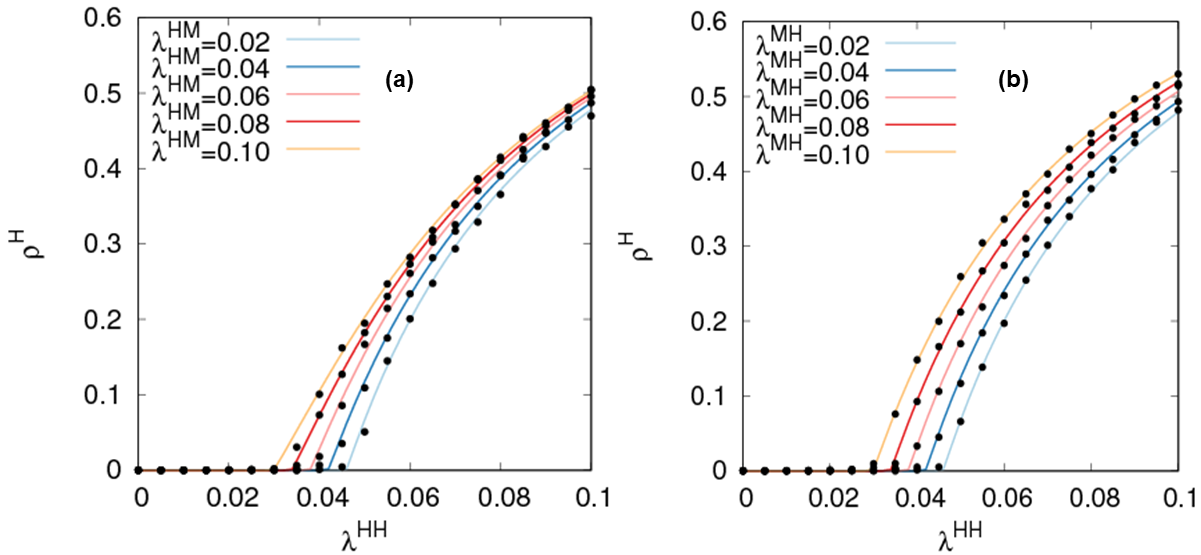}
\caption{Fraction of infected people in the stationary state $\rho^H$ as a function of the human-to-human contagion probability $\lambda^{HH}$. Solid lines correspond to the solution obtained by iterating Eqs.~(\ref{ec.human})-(\ref{ec.mosquito}) whereas black dots represent the solutions from Monte Carlo simulations obtained after averaging over $20$ realizations. In panel (a) the color of each line denotes the vector-human contagion rate $\lambda^{HM}$, while the rest of parameters of our model ($\lambda^{MH},\beta,k,\gamma,\mu^H,\mu^M$) are set to ($0.01,2,4,2,0.2,0.1$). In Panel (b) the color of each line denotes the vector-human contagion rate $\lambda^{MH}$, while the rest of parameters of our model ($\lambda^{HM},\beta,k,\gamma,\mu^H,\mu^M$) are set to ($0.01,2,4,2,0.2,0.1$).}
\end{figure}

Once analyzed the role that vector biting rate, $\beta$, plays on epidemic spreading, it is worth analyzing the effect of breaking the symmetry between the contagion rates between humans and vectors, $\lambda^{MH}=\lambda^{HM}$, assumed in Fig.~2.a. For this purpose, in Fig.~3.a we have analyzed the disease incidence as a function of $\lambda^{HH}$ for several values of $\lambda^{HM}$, fixing the biting rate to $\beta=2$ and the vector-human contagion probability to $\lambda^{MH} = 0.01$. Alternatively, in Fig.~3.b the same analysis is performed by fixing $\lambda^{HM} = 0.01$ and exploring different values of $\lambda^{MH}$. Although in both cases the epidemic threshold, $\lambda_c^{HH}$, decreases as the likelihood of crossed transmission [humans to vectors in (a) and vectors to humans in (b)], it is remarkable that this trend is much smoother than the observed in Fig.~2.a when increasing the biting rate. Given the observed agreement between Monte Carlo and mean-field equations in Fig.~2.a and Fig.~3, we can explain this result from the different functional roles played by $\beta$ (exponent) and the crossed contagion probabilities in Eqs.~(\ref{ec.human})-(\ref{ec.mosquito}).

\subsection{Epidemic threshold}
\label{subsec.threshold}

Once the validity of mean-field Eqs.~(\ref{ec.human})-(\ref{ec.mosquito}) has been shown, we  can now make use of them and obtain an analytical expression for the conditions to be fulfilled at the epidemic onset. It is worth mentioning that in the original RM model the epidemic solution appears when \cite{Rohani}:
\begin{equation}
\frac{\beta^2\gamma\lambda^{HM}\lambda^{MH}}{\mu^{M}\mu^{H}}=1\;.
\label{RMthreshold}
\end{equation} 
From the former expression one can derive different epidemic thresholds, {\em i.e.} those critical values for the RM parameters for which the l.h.s. of Eq.~(\ref{RMthreshold}) is equal to 1.

Coming back to our generalized RM model, Eqs.~(\ref{ec.human})-(\ref{ec.mosquito}), let us first assume that we are under stationary conditions, {\em i.e.} the disease has reached the stationary solution: $\rho^H(t+1) = \rho^H(t) = \rho^H$ and $\rho^M(t+1) = \rho^M(t) = \rho^M$. Introducing these two stationary values, $\rho^H$ and $\rho^M$, into Eqs.~(\ref{ec.human})-(\ref{ec.mosquito}) yields:
\begin{eqnarray}
\mu^M\rho^M &=& \left(1-\rho^M\right)\left(1-\left(1-\lambda^{HM}\rho^{H}\right)^\beta\right)\nonumber \label{Ec3}\\
\mu^H\rho^H &=& \left(1-\rho^H\right)\left[1-\left(1-\lambda^{MH}\rho^{M}\right)^{\beta\gamma}\left(1-\lambda^{HH}\rho^{H}\right)^{k}\right]\nonumber .
\label{Ec4} 
\end{eqnarray}
Close enough to the epidemic onset, we can suppose that $\rho^H$ and $\rho^M$ are small enough ($\rho^H=\epsilon^{H}$ and $\rho^M=\epsilon^{M}$) allowing us to linearize the former equations:
\begin{eqnarray}
\mu^H\epsilon^H &=&  k\lambda^{HH} \epsilon^H + \beta\gamma\lambda^{MH}\epsilon^M \label{rhoh}\ ,\\
\mu^M\epsilon^M &=& \beta\lambda^{HM}\epsilon^H \label{rhom}\ .
\end{eqnarray}
By substituting Eq. \ref{rhom} into Eq. \ref{rhoh} and rearranging the terms, we obtain the following equation:
\begin{equation}
\epsilon^H\left(\mu^H - k\lambda^{HH} - \frac{\beta^{2}\gamma\lambda^{HM}\lambda^{MH}}{\mu^M}\right)=0.
\end{equation}
Since we have considered $\epsilon^H$ as a negligible but non-zero, the right term must be equal to zero and the new condition for the epidemic onset reads:
\begin{equation}
\frac{k\lambda^{HH}}{\mu^{H}} + \frac{\beta^{2}\gamma\lambda^{HM}\lambda^{MH}}{\mu^M\mu^{H}}=1\;.
\label{newRMthreshold}
\end{equation}
Comparing Eq.~(\ref{newRMthreshold}) with Eq.~(\ref{RMthreshold}), we easily notice the correction $k\lambda^{HH}/\mu^{H}$ provided by the addition of  human-to-human contagions. It is remarkable that $k\lambda^{HH}/\mu^{H}=1$ corresponds to the epidemic threshold in a SIS model where each human contacts with other $k$ agents and contagions take place exclusively by human-to-human interactions.

Considering now the probability of contagion between humans as our control parameters (as in Fig.~2.a and Fig.~3), the epidemic threshold, $\lambda_{c}^{HH}$, can be estimated from Eq.~(\ref{newRMthreshold}) as:
\begin{equation}
\lambda_{cr}^{HH} = \frac{\mu^H}{k} - \frac{\beta^2\gamma\lambda^{HM}\lambda^{MH}}{k\mu^M}\ .
\label{Threshold}
\end{equation}
Interestingly, Eq. (\ref{Threshold}) can give rise to negative values of the epidemic threshold, which in physical terms correspond to the situation in which  the vector-human contagion path is strong enough to sustain the epidemic solution, and for any value of $\lambda^{HH}$ the epidemic solution is the equilibrium point. Therefore, the epidemic threshold $\lambda_c^{HH}$ will be that predicted by Eq. (\ref{Threshold}), unless the l.h.s of Eq. (\ref{Threshold}) is negative so that $\lambda_c^{HH}=0$. In Fig.~4 we show the excellent agreement between our prediction for $\lambda_c^{HH}$ (black solid line) and its value according to Monte Carlo simulations by plotting the fraction of infected humans as a function of $\beta$ and $\lambda^{HH}$. Apart from the agreement, it is remarkable that an increase of the biting rate $\beta$ leads to a decrease of the epidemic threshold until it vanishes for $\beta =10$.

\begin{figure}[t!]
\centering
\includegraphics[scale=0.35,angle=-90]{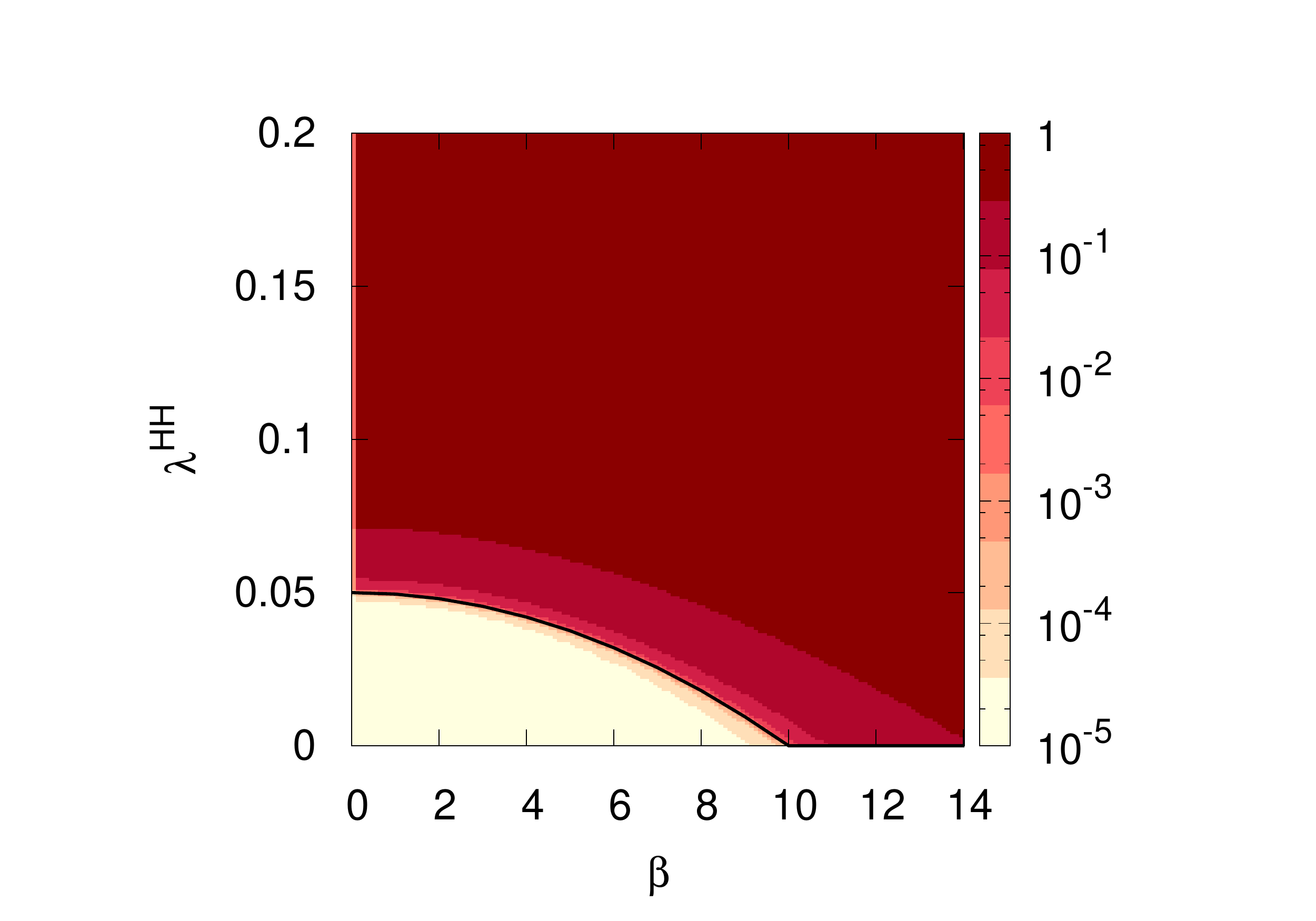}
\caption{Fraction of infected people in the stationary state denoted by the color scale for each value of ($\beta,\lambda^{HH}$). Solid black line correspond to the estimation of the epidemic threshold as a function of $\beta$ using Eq. (\ref{Threshold}). The epidemic parameters involved in contagion processes are set to ($\lambda^{MH},\lambda^{HM},k,\gamma$) = ($0.01,0.01,4,2$). Recovery rates are set to ($\mu^H,\mu^M$) $=$ ($0.2,0.1$). }
\end{figure}

\section{Contacts network approach}
In order to add more realism to the generalization of the RM model, we now leave the mean field hypothesis that assumes agents to be homogenous and statistically equivalent. To this aim, we now consider that contacts among humans are described by a complex network represented by a matrix $\textbf{A}$, whose elements determine the interaction between human. This way, $A_{ij}=1$ when agents $i$ and $j$ interact while $A_{ij}=0$ otherwise. On the other hand, we keep considering the mean-field dynamics for the set of vectors.

In this framework Eqs.~(\ref{ec.human})-(\ref{ec.mosquito}) are no longer valid since we need $N$ equations to characterize the evolution of the state of each agent and another one for the evolution of the vector population. Under these premises, the probability than an agent $i$ is infected at time $t+1$ is given by:
\begin{widetext}
\begin{equation}
\rho_i^{H} (t+1) = \left(1-\mu^H\right)\rho_i^H(t) + \left(1-\rho_i^H(t)\right)\left[1-\left(1-\lambda^{MH}\rho^{M}\right)^{\beta\gamma}\prod\limits_{j=1}^{N}\left(1-A_{ij}\lambda^{HH}\rho_j^{H}\right)\right]\ .
\label{rhoh2}
\end{equation}
\end{widetext}
On the other hand, the equation for the fraction of infected vectors reads as in the mean-field case, Eq.~(\ref{ec.mosquito}):
\begin{widetext}
\begin{equation}
\rho^M(t+1) = \left(1-\mu^M\right)\rho^M(t) \nonumber+ \left(1-\rho^M(t)\right)\left(1-\left(1-\lambda^{HM}\rho^H(t)\right)^\beta\right) \ ,
 \label{rhom2}
\end{equation}
\end{widetext}
with the exception that here $\rho^H(t)$, which denotes the total fraction of infected people at time $t$, is calculated as the average of the set of $N$ probabilities $\{\rho_i^{H} \}$ as:
\begin{equation}
\rho^H (t) = \frac{1}{N}\sum\limits_{j=1}^N\rho_i^H(t)\ .
\end{equation}

\begin{figure}[b!]
\centering
\includegraphics[scale=0.3,angle=-90]{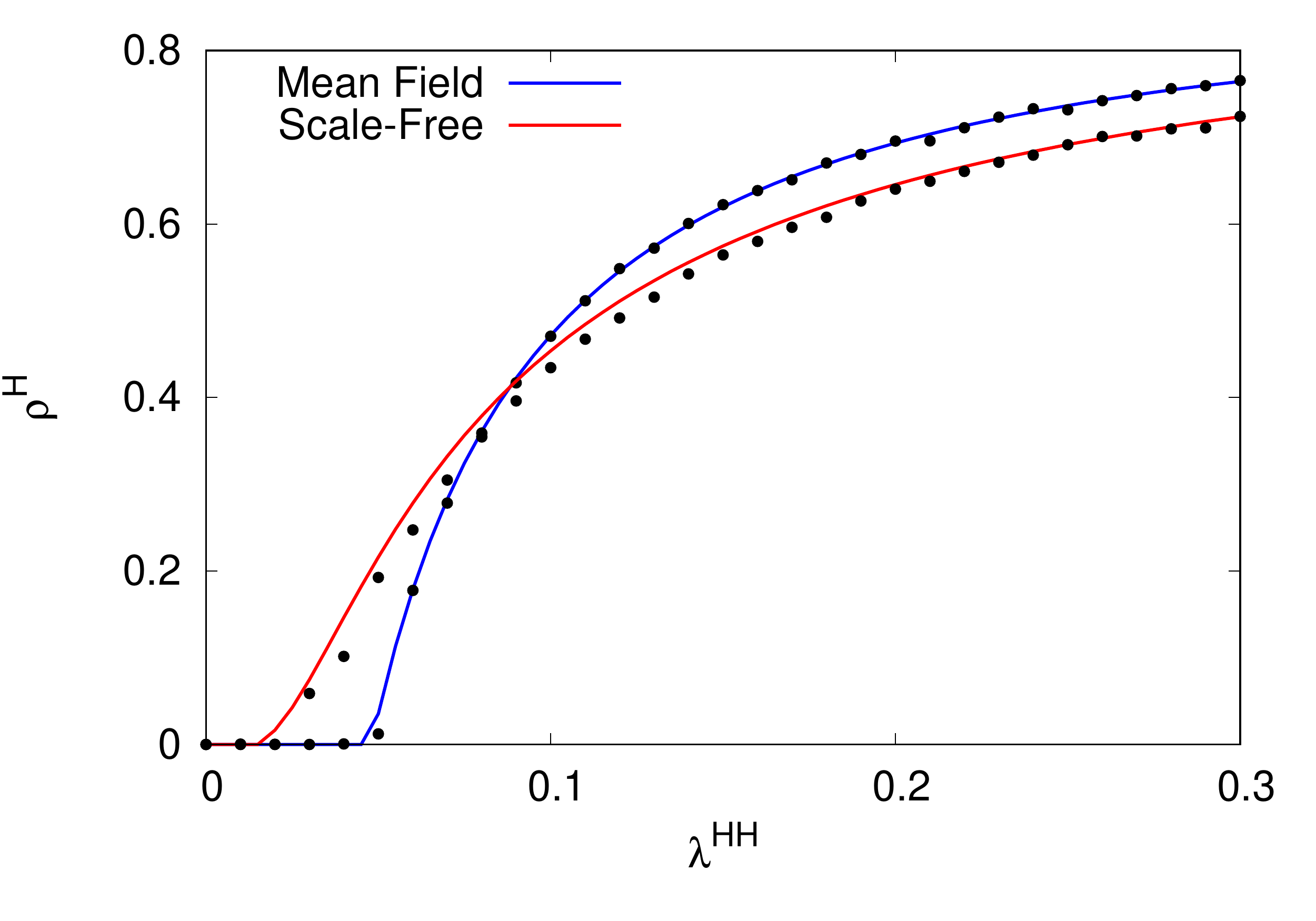}
\caption{Fraction of infected people in the stationary state $\rho^H$ as a function of the human-to-human contagion probability $\lambda^{HH}$. The contagion parameters are set to ($\lambda^{MH},\lambda^{HM},\beta,\gamma$) = ($0.01,0.01,2,2$). Solid lines show the predictions of the proposed models, encoding the human contacts by a mean field theory (blue) with  $k=4$ or by a scale-free network with $\langle k\rangle = 4$. Black dots correspond to the results from MC simulations for each case, both obtained by averaging $20$ realisations. The recovery rates are set to $(\mu^H,\mu^M)= (0.2,0.1)$.}
\end{figure}

In order to validate these equations, we assume that contact among humans are determined by scale-free (SF) networks. The main characteristic of these networks is the heterogeneity for the number of contacts per individual, since they are composed by a large number of nodes with low connectivity and also by a few highly connected  ones that are the so-called hubs. In our case the 
SF network is constructed following the Barab\'asi-Albert method \cite{BA} that generates graphs for which the probability of finding a node with $k$ contacts is given by $P(k)\sim k^{-3}$. As in the case of the mean-field formulation, we compare the predictions of Eqs. (\ref{rhoh2})-(\ref{rhom2}) with the results obtained from Monte Carlo simulations. These simulations are carried out in a similar way to those of the mean field case, but taking into account that human contacts are now governed by a SF contact network. \\

In Fig. 5 we show the epidemic diagram $\rho^{H}(\lambda^{HH})$ for the SF network and that of a mean-field population, in both cases the average number of contacts if set to $\langle k\rangle =4$. Again the solid lines show the solution obtained by iterating Eqs. (\ref{rhoh2})-(\ref{rhom2}) and dots correspond to the results from Monte Carlo simulations. The agreement between theory and simulations is still quite good. Besides, we appreciate how introducing heterogeneity in terms of human contacts leads to an important decrease of the epidemic threshold due to the presence of hubs that, due to their large number of acquaintances, boost the propagation of the disease.

\subsection{Epidemic threshold with contact networks}

To round off, we tackle the derivation of an analytical expression for the epidemic threshold that allows us to analyze the role of the degree heterogeneity of the human contact network. For this purpose, we proceed as in the mean-field case and consider stationarity and a small disease incidence. This way, the probability that an agent $i$ is infected  $\epsilon_i^H$ as well as the fraction of infected vectors are very small, so that $\rho_i^H=\epsilon_i^H\ll 1 \  \forall i$ and $\rho^M = \epsilon^M \ll 1$. This approximation allows us to linearize Eqs. (\ref{rhoh2})-(\ref{rhom2}) yielding the following set of equations:
\begin{eqnarray}
\mu^H\epsilon_i^H &=& \sum\limits_{j=1}^N \lambda^{HH} A_{ij}\epsilon_j^H + \beta\gamma\lambda^{MH}\epsilon^M\label{eqhlin} \ ,\\
\mu^M\epsilon^M &=& \beta\lambda^{HM}\sum\limits_{j=1}^N\frac{\epsilon_j^H}{N}\label{eqmlin}\ .
\end{eqnarray}
To reduce the complexity of this set of equations, we introduce the solution for $\epsilon^M$, Eq.(\ref{eqmlin}), into the first $N$ ones, Eq.~(\ref{eqhlin}), which now read as:
\begin{equation}
\mu^H\epsilon_i^H = \sum\limits_{j=1}^N\left(\lambda^{HH} A_{ij} + \frac{\beta^2\gamma\lambda^{MH}\lambda^{HM}}{N\mu^M}\right)\epsilon_j^H\ .
\end{equation}
At this point, to get a compact expression of the epidemic threshold, $\lambda_c^{HH}$, let us express, without any loss of generality, the cross-contagion rates $\lambda_c^{HM}$ and $\lambda_c^{MH}$ as a function of the human-human one $\lambda^{HH}$. In particular, we define $\lambda^{MH}=\alpha^{MH}\lambda^{HH}$ and $\lambda^{HM} = \alpha^{HM}\lambda^{HH}$. This enables us to write: 
\begin{equation}
\frac{\mu^H}{\lambda^{HH}}\epsilon_i^{H} =  \left(\textbf{M}{\boldsymbol\epsilon}\right)_i\ ,
\label{eigs}
\end{equation}
where {\bf M} is a $N\times N$ matrix whose elements are given by:
\begin{equation}
{\bf M}_{ij} = A_{ij} + \frac{\beta^2\gamma\alpha^{MH}\alpha^{HM}}{N\mu^M}
\label{matrixM}\ .
\end{equation}
Equation (\ref{eigs}) encodes an eigenvalue problem, and there are $N$ solutions of $\lambda^{HH}$, each one associated with one eigenvalue of matrix ${\bf M}$. However, since we are interested in the minimum probability of contagion for which the epidemic solution exists, the epidemic threshold is given by:
\begin{equation}
\lambda_c^{HH}=\frac{\mu^H}{\Lambda_{\text{max}}(\textbf{M})} \ ,
\end{equation}
where $\Lambda_{\text{max}}(\textbf{M})$ denotes the maximum eigenvalue of $\textbf{M}$. 

Interestingly, for the case of a SIS disease in contact networks, the epidemic threshold is proportional to the inverse of the maximum eigenvalue of the Adjacency matrix, $\lambda_c\sim1/\Lambda_{\text{max}}(\textbf{A})$, whose value increases with degree heterogeneity. Here, this threshold is modified as matrix ${\bf M}$, Eq. (\ref{matrixM}), is the sum of ${\bf A}$ plus a positive constant. Thus, the vector-human contagion path decreases $\lambda_c^{HH}$ with respect to that expected from simple human-to-human infections.

\section{Conclusions}

The great concern raised by the rapid spread of ZIKV pathologies has spurred the scientific research about the particular features that characterize this disease. As an example, A. Allard et al \cite{Allard1,Allard2} have shown, by using bond percolation, that the well-known contagion asymmetry between males and females \cite{PLOS7,PLOS8} leads to the apparition of a double epidemic threshold: one associated to the onset of epidemics inside the men-who-have-sex-with-men (MSM) community and another one associated to its global outbreak. Interestingly, this high asymmetry also allows the disease to be self-sustained, even in the case that the vector-human reproductive ratio is small. This great impact of the human-human contagion path has already been reported by other mean field theories \cite{PLOS,S.Rep1} that were proposed to reproduce ZIKV contagion mechanisms.

In this paper, we have formulated a new version of the RM model in order to incorporate a feature that differentiates ZIKV transmission from most of the usual VBD: the presence of human-to-human contagions. We have first proposed a mean field theory which enables us to get some intuition about the effect of adding this new contagion path. To validate this theory, we have compared its predictions about the incidence as well as the temporal evolution of a disease to results from numerical Monte Carlo simulations. Moreover, we have observed how introducing a second contagion path boost the spreading of a VBD since the epidemic solution can be found despite being below the epidemic threshold of the original RM model. In this sense, to completely characterize the influence of human contacts on the propagation of this kind of diseases, we have linearized the mean field equations, obtaining the lowest value of the human-human contagion rate which leads to a non-zero impact of a VBD with human interactions.\\

To gain further insight about the importance of human-to-human contacts we have included the existence of a human contact network that acts as the backbone of human-to-human infections. Specifically, we have observed how degree heterogenous contact topologies, like scale-free networks, makes the population prevention from this kind of diseases much more difficult, since the epidemic threshold decreases considerably with respect to the mean-field theory. In this regard, we have been able to deduce an analytical expression for the epidemic threshold capturing both the influence of the network of human contacts and the usual parameters of the RM model.\\

Although our model is a theoretical approach which is not designed to tackle the specific case of ZIKV transmission, the introduction of a microscopic framework characterizing human-to-human contagions into the original RM model will pave the way to the formulation of more accurate ZIKV models. In particular, our formalism can be the starting point to elaborate future metapopulation models \cite{metapop1,metapop2,metapop3,metapop4} in which more realistic ingredients such as human mobility or seasonal fluctuations of vector contagion rates \cite{seasonal1,seasonal2}, are incorporated.

\section*{Acknowledgements}
%%%%%%%%%%%%%%%%%%%%%%%%%%%%%%%%%%%%%%%%%%%%%
DSP and JGG acknowledge financial support from MINECO (projects FIS2014-55867-P and FIS2017-87519-P) and from the Departamento de Industria e Innovaci\'on del Gobierno de Arag\'on y Fondo Social Europeo (FENOL group E-19). FNM acknowledges support from the DIN-UPTC under project SGI 1958.
.


\begin{thebibliography}{}
\bibitem{WHO}
WHO Factsheet {\bf 38} (2014)
\bibitem{climate}
E. A. Gould, S. Higgs, Trans R Soc Trop Med Hyg {\bf 103} (2), 109 (2009)
\bibitem{N.climate}
P. W. Gething et al, Nature {\bf 465}, 342 (2010)
\bibitem{Sharp}
T. M. Sharp, K. M. Tomashek, Curr. Epi. Reports {\bf 4} (1), 11 (2017)
\bibitem{Albo_global}
ECDPC. {\it Development of Aedes albopictus risk maps} (2009)
\bibitem{Brown} J. E. Brown, E. J. Scholte, M. Dik, H. W. Den, J. Beeuwkes, and J. R. Powell, Emerg Infect Dis. 17, 2335 (2011).
%Aedes aegypti mosquitoes imported into the Netherlands, 2010. ; 17: 2335-2337
\bibitem{Romi} R. Romi, F. Severini and L. Toma, J Am Mosq Control Assoc. 22, 149 (2006)
%Cold acclimation and overwintering of female Aedes albopictus in Roma. .
\bibitem{Madeira}
A. Wilder-Smith et al, Euro Surveill. {\bf 19} (8), 20718 (2014)
\bibitem{Demir}
T. Demir, S. Kilic, Folia Microbiologia {\bf 61} (6), 523 (2016)
\bibitem{Zika}
H. Sakkas, V. Economou, C. Papadopoulou, Journal of vector borne diseases {\bf 53} (4), 305 (2016)
\bibitem{Zika2}
World Health Organization, Zika virus, microcephaly and Guillain-Barr\'e syndrome situation report (2016)
\bibitem{Zikah}
F. Grischott, M. Puhan, C. Hatz, P. Schlagenhauf, Travel Med Infect Dis.  {\bf 14} (4), 313 (2016) 
\bibitem{Anderson}
R. M. Anderson, R.L. May, {\it Infectious Diseases of Humans: Dynamics and Control} (Oxford University Press, Oxford, 1991)
\bibitem{Rohani}
J. M. Keeling, P. Rohani, {\it Modeling Infectious Diseases in Humans and Animals} (Princeton University Press, Princeton, 2007)
\bibitem{Bacaer2011}
N.A. Baca\"er, {\it A Short History of Mathematical Population Dynamics} (Springer, London, 2011)
\bibitem{Ross}
R. Ross, {\it The prevention of Malaria} (John Murray, London, 1911)
\bibitem{Macdonald}
G. Macdonald,  {\it The Epidemiology and Control of Malaria} (Oxford University Press, Oxford, 1957)
\bibitem{N.education}
K. Magori, J. M. Drake, Nat. Ed. Knowledge {\bf 4} (4), 14 (2013)
\bibitem{Smith}
D. L. Smith et al, PLOS Pathogens {\bf 8} (4), 1 (2012)
\bibitem{Ruan2008}
S. Ruan, D. Xiao, J. C. Beier, Bulletin of Mathematical Biology {\bf 70}, 1098 (2008)
\bibitem{Regan2016}
S. M. O'Regan, J. W. Lillie, J. M. Drake, Theoretical ecology {\bf 9} (3), 269 (2016)
\bibitem{review}
R. C. Reiner et al, J. Roy. Soc. Interface {\bf 10(81)}, 20120921 (2013).
\bibitem{Newman}
M. E. J. Newman, SIAM review {\bf 45} (2), 167 (2001)
\bibitem{Boccaletti}
S.Boccaletti et al, Physics Reports {\bf 424} (4), 175 (2006)
\bibitem{MMCA1}
S. G\'omez et al, Europhys. Lett. {\bf 89}, 38009 (2010)
\bibitem{MMCA2}
B. Guerra and J. G\'omez-Garde\~nes, Phys. Rev. E {\bf 82}, 035101(2010).
\bibitem{MMCA3}
S. G\'omez, J. G\'omez-Garde\~nes, Y. Moreno, A. Arenas, Phys. Rev. E {\bf 84}, 036105 (2011).
\bibitem{RMP}
R. Pastor-Satorras C. Castellano, P. Van Mieghem, A. Vespignani, Rev. Mod. Phys. {\bf 87}, 925 (2015)
\bibitem{BA}
A. L. Barab\'asi, R. Albert, Science {\bf 286} (5439), 509 (1999)
\bibitem{Newman2}
M. E. J. Newman, {\it Networks: An Introduction} (Oxford University Press, Oxford 2010)
\bibitem{socialnet}
E. Cho, S. A. Myers and J. Leskovec. \textit{Proceedings of the 17th ACM SIGKDD international conference on Knowledge discovery and data mining.} (ACM, 2011) p. 1082-1090
\bibitem{Allard1}
A. Allard, B. M. Althouse, S. V. Scarpino and L. H\'ebert-Dufresne, PNAS {\bf 114}, 8969 (2017)
\bibitem{Allard2}
A. Allard, B. M. Althouse, L. H\'ebert-Dufresne and S. V. Scarpino, PLOS Pathogens {\bf 13}, e1006633 (2017)
\bibitem{PLOS7}
B. Visseaux et al, Lancet Infect Dis. \textbf{16}, 1000–01 (2016)
\bibitem{PLOS8}
E. Nicastri et al, Eurosurveillance {\bf 21}, 32 (2016). 
\bibitem{PLOS}
A. J. Kucharski et al, PLoS Negl Trop Dis. {\bf 10} (5), e0004726 (2016)
\bibitem{S.Rep1}
D. Gao et al, Scientific Reports {\bf 6}, 28070 (2016)
\bibitem{metapop1}
J. G\'omez-Garde\~nes, D. Soriano-Pa\~nos and A. Arenas, Nat. Phys, doi:
10.1038/s41567-017-0022-7 (2018).
\bibitem{metapop2}
V. Colizza, R. Pastor-Satorras and A. Vespignani, Nat. Phys \textbf{3}, 276 (2007) 
\bibitem{metapop3}
V. Belik, T. Geisel and D. Brockmann, Phys. Rev. X \textbf{1}, 011001 (2011)
\bibitem{metapop4}
J. T. Matamalas, M. De Domenico and A. Arenas, J. R. Soc. Interface \textbf{13}, 20160203 (2016)
\bibitem{seasonal1}
S. Polwiang, PeerJ \textbf{3}, e1069 (2015)
\bibitem{seasonal2}
J. Rockl\"ov et al, EBioMedicine \textbf{9}, 250 (2016)
\end{thebibliography}
\end{document}